\documentclass[twocolumn,showpacs,preprintnumbers,amsmath,amssymb,showkeys,nofootinbib]{revtex4}

\usepackage{dcolumn}
\usepackage{bm}

\usepackage[T1]{fontenc}
\usepackage{ae,aecompl}
\newcommand{\be}{\begin{equation}}
\newcommand{\ee}{\end{equation}}

\usepackage{graphicx}	
\usepackage{amsmath}	
\usepackage{amssymb}	

\usepackage{xcolor} 
\usepackage{cancel} 
\usepackage{ulem}

\begin{document}

\title{On self-gravitating strange dark matter halos around galaxies}
\author{Marco Merafina}
\email{marco.merafina@roma1.infn.it}
\affiliation{Department of Physics, University of Rome La Sapienza, Rome, Italy}
\affiliation{INFN Laboratori Nazionali di Frascati, Frascati, Rome, Italy}
\author{Francesco G. Saturni}
\email{francesco.saturni@inaf.it}
\affiliation{INAF Osservatorio Astronomico di Roma, Monte Porzio Catone, Rome, Italy}
\affiliation{ASI Space Science Data Center, Rome, Italy}
\author{Catalina Curceanu}
\email{catalina.curceanu@lnf.infn.it}
\affiliation{INFN Laboratori Nazionali di Frascati, Frascati, Rome, Italy}
\author{Raffaele Del Grande}
\email{raffaele.delgrande@lnf.infn.it}
\affiliation{INFN Laboratori Nazionali di Frascati, Frascati, Rome, Italy}
\affiliation{Centro Fermi - Museo Storico della Fisica e Centro Studi e Ricerche ``Enrico Fermi'', Roma, Italy}
\author{Kristian Piscicchia}
\email{kristian.piscicchia@cref.it}
\affiliation{Centro Fermi - Museo Storico della Fisica e Centro Studi e Ricerche ``Enrico Fermi'', Roma, Italy}
\affiliation{INFN Laboratori Nazionali di Frascati, Frascati, Rome, Italy}

\begin{abstract}
A new family of nonrelativistic, Newtonian, non-quantum equilibrium configurations describing galactic halos is introduced, by considering strange quark matter conglomerates with masses larger than about 8\,GeV as new possible components of the dark matter. Originally introduced to explain the state of matter in neutron stars, such conglomerates may also form in the high-density and temperature conditions of the primordial Universe and then decouple from ordinary baryonic matter, providing the fundamental components of dark matter for the formation of pristine gravitational potential wells and the subsequent evolution of cosmic structures. The obtained results for halo mass and radius are consistent with the rotational velocity curve observed in the Galaxy. Additionally, the average density of such dark matter halos is similar to that derived for halos of dwarf spheroidal galaxies, which can therefore be interpreted as downscaled versions of larger dark matter distributions around Milky Way-sized galaxies and hint for a common origin of the two families of cosmic structures.
\end{abstract}


\maketitle



\section{Introduction}

Dark matter (DM) is one of the current challenges for modern astrophysics. Originally introduced to explain the flat rotation curves of spiral galaxies \cite{1}, DM is also required as a fundamental component ($\sim 30\%$; see e.g.\,\cite{2} and references therein) of the Universe's energy content. Unfortunately, the DM weird physical property to not couple with radiation as the baryonic matter, required in order to explain its invisibility to traditional astronomical observations (e.g.\,\cite{3}), prevents astrophysicists to directly provide data on its constituents.

Several efforts have been made in order to identify plausible DM candidates, both as elementary particles (see e.g.\,\cite{4} for a review) and macroscopic objects (MACHOs; e.g.\,\cite{5}). However, without any direct hint about DM physics, the parameter space covered by the families of plausible DM candidates extends over many orders of magnitude of masses (e.g. the elementary particles, ranging from $\sim 10^{-15}$\,GeV of axions up to $\sim 10^{15}$\,GeV of ``wimpzillas'') and cross sections (from $\sim 10^{-35}$\,pb of gravitinos up to $\sim 1$\,pb of neutrinos).

In past years, a common idea of DM physics was that it is related to families of particles beyond the Standard Model (SM), such as those arising from supersymmetric theories (e.g.\,\cite{6}). However, extensive runs performed at the Large Hadron Collider (LHC), in order to unveil signatures of events with production of non-interacting DM, have shown no clear features of such phenomena in large energy ranges, thus leading to a progressive exclusion of supersymmetric DM particles (e.g.\,\cite{7}) although some possibilities still remain (e.g.\,\cite{8}). Similarly, the investigation of the density of MACHOs in the Milky Way and in extragalactic halos has shown that such objects are not abundant enough to represent a significant fraction of the DM mass (e.g.\,\cite{9}).

The possibility that DM is composed by conglomerates of matter containing roughly the same number of up, down and strange quarks is also challenging. The properties and the stability of such strange quark matter (SQM), which were conjectured long time ago (see e.g.\,\cite{62,63}), are still debated in literature. Although the formation process of SQM conglomerates is astrophysically disadvantaged, favourable conditions for their formation could have been achieved in the early Universe, soon after the Big Bang. The ALICE experiment at LHC has recently measured a contribution to the cosmic ray flux which is compatible with a SQM component \cite{65}; further investigations on this item are then required. To this end, despite the problem of the stability with respect to the strong, weak and electromagnetic interactions, it is worth investigating the gravitational properties of galactic halos composed by SQM conglomerates, not explored so far.

Recently, a framework for (indirect) search of DM signals also arises in astrophysics from the possibility that DM particles \textit{self}-interact via annihilation or decay to produce SM pairs \cite{73}, that subsequently annihilate into final-state photons (e.g.\,\cite{10,11}). Such photons could be indirectly detected on Earth via the emissions produced by Cherenkov radiation in the atmosphere if the mass of the pristine DM particles is sufficiently high (more than some GeV). The task of discovering self-interacting DM through Cherenkov observations is potentially at reach of next-generation Cherenkov telescope (e.g.\,\cite{12}).

In this paper, the self-gravitational equilibrium of SQM galactic halos in Milky Way-sized spiral galaxies is calculated and the corresponding gravitational properties are constrained for the first time. Conglomerates of SQM with mass larger than about 8\,GeV are considered as components of the DM, nevertheless the chosen nature for the DM particle is found to not affect the validity of the obtained results which depend only on the mass.

We then compare the average properties of such a halo with the corresponding quantities derived from the spherical Monte-Carlo Jeans analysis of a sample of DM halos around dwarf spheroidal galaxies (e.g.\,\cite{97}).

\section{Massive particles as dark matter candidates}

Besides the most commonly investigated DM candidates -- e.g. Weakly-Interacting Massive Particles (WIMPs), or axions -- , for which no compelling observational or experimental evidence still exists, nor any proved beyond-the-Standard-Model production mechanism, other DM candidates were theorized (ranging from heavy stable particles to new states of matter) including ones which might arise within in the SM.

If DM particles have no asymmetry and self-annihilate with a cross section comparable to the electroweak scale, then the expected amount of DM in the Universe can be achieved provided that the particle mass ranges within the GeV and the TeV scales. Such particles are known as WIMPs (e.g.\,\cite{36a,36b}). Alternatively, if DM owns a particle-antiparticle asymmetry in any conserved quantum number, as it is for baryons, then the correct relic abundance should be accomplished in a completely different way, which may be related to the asymmetry in the baryonic sector \cite{36c}.

A broad zoology of new Stable and Massive Particles (SMPs) has been proposed based on general considerations on cosmology and DM, which include heavy leptons and hadrons, fractionally charged particles, mirror matter and SQM. Further candidates were postulated which could arise as topological field configurations like magnetic monopoles, Q-balls or black holes. A complete review about the scenarios predicting SMPs is given in \cite{37}, with particular focus on the experimental, non-collider, techniques used for their search.

In the supersymmetric (SUSY) extension of the SM stable particles involving heavy leptons and hadrons can arise as possible DM candidates, with the hypotised properties of SMPs. Stable negatively charged particles (X) could bind with nuclei to form dense neutral objects which could act as DM \cite{38,39}, for instance X$^-$p and X$^{--}$He. It was argued that DM species such as X$^{--}$He could provide an explanation to the observed annual modulation of the ionization signal in the DAMA/NaI and DAMA/LIBRA experiments \cite{40} but the scarce abundance of anomalous heavy isotopes in terrestrial matter strongly constraints the models. The searches at LHC for stable interacting particles exclude masses up to $\sim 1$\,TeV, whereas the limit on the lepton-like objects masses is of $\sim 300$\,GeV \cite{41,42,43,44,45,46,47}.

Other cosmological scenarios involve fractionally charged particles forming composite objects which could, in principle, be DM candidates \cite{48}. The charge quantization arises in fundamental theories and it does not guarantee that charges have values that are integer multiples of the elementary charge $e$. In principle fractionally charged particles could exist without conflicts with the theory \cite{49,50}, nevertheless there is no experimental evidence of such exotic states inside the SM. Composite objects of fractionally charged particles, if they exist, are strongly constrained by cosmological arguments to have large masses (of the order of 10$^{12}$\,GeV).

The possibility that DM is composed of dark particles, which are decoupled from the ordinary particles unless additional interactions are assumed, has been studied extensively in the last few decades \cite{51,52,53,54,55,56,57,58,59,60,Va1,
Va2}. The simplest of such models assumes that the matter of the dark sector, known as ``mirror matter", is coupled to ordinary matter through the kinetic mixing of the dark and ordinary electromagnetism and predicts the existence of dark atoms. Although its strong self-interaction cross section would not qualify the ``mirror matter" as a good candidate for the DM, the possible existence of dark objects gravitationally bound to the Solar System could help to explain the anomalies in the behaviour of some meteoroids \cite{61}.

The stability of SQM conglomerates, containing roughly the same numbers of $u$, $d$ and $s$ quarks, was conjectured long time ago (see e.g.\,\cite{62,63}). It is argued that, despite the big mass of the $s$ quark, compared to the $u$ and $d$ quarks, this is smaller than the chemical potential due to the Pauli exclusion principle in bulk quark matter, making such a mixture energetically favoured. The only way for standard baryonic matter to make the transition to an SQM phase would be $u$ and $d$ quarks conversions into $s$ quarks via weak interactions, stabilized by the chemical potential release. Such a process is disadvantaged in stellar nuclear reactions, whereas the SQM lumps formation could have found favourable conditions in the early Universe. In particular, if the Quantum Chromodynamics (QCD) phase transition is first order, the dynamics of bubble nucleation are such that quark matter lumps would form at that stage and shrink and cool, moving on the Equation-of-State diagram from a high temperature to a zero temperature, high chemical potential configuration. 

Stable SQM lumps were conjectured with baryon numbers $A$ ranging from few unities to $10^{57}$ \cite{64}, limit for which the \textit{strange star} would collapse into a black hole.

In \cite{65} a recent Cosmic Ray (CR) measurement, performed by the ALICE experiment at CERN LHC in its dedicated CR run \cite{66,67}, is investigated. In \cite{67} high multiplicity muon bundles were detected in extensive air showers produced by the CR interactions in the upper atmosphere. The analysis described in \cite{65} is focused on those events which contain more than 100 reconstructed muons. This in motivated by the fact that Monte Carlo simulations assuming a standard combination of proton and iron components in the primary CR flux are not suitable to describe the higher multiplicity events (with more than 100 muons). It is shown that the ALICE measurement is compatible with an SQM component in the CR, characterised by a very high baryon number of the order of $A\sim 10^3$ (according to \cite{68} exceeding a critical value of $A\sim 300 \div 400$ the SQM lumps are absolutely stable against neutron emission, below this limit they rapidly decay by evaporating neutrons). The frequency distribution of the highest multiplicity events could be reproduced assuming an SQM abundance in the primary CR flux of the order of $10^{-5}$ of the same total energy per particle, value which seems to be consistent with all the recent observations (e.g.\,\cite{69}).  

In \cite{70} it is argued that stable bumps of matter with strange quark content could be obtained in 
$\Lambda(1405)$ conglomerates, whose formation may be conceived during the Big-Bang Quark Gluon Plasma (QGP) period in the early universe. The argument exposed in \cite{70} is based on the attractive isospin $I=0$ antikaon-nucleon ($\mathrm{\bar{K}}$N) strong interaction at energies below the $\mathrm{\bar{K}}$N mass threshold, which appears to be strong enough to form a $\mathrm{\bar{K}}$N bound state. In \cite{70} the $\Lambda$(1405) resonance (also indicated with $\Lambda^*$), which according to the chiral models (see e.g.\,\cite{IkedaWeise,Cieply,Guo,Mai,Feijoo}) emerges just few MeV below the $\mathrm{\bar{K}}$N threshold, is instead interpreted as a ($\mathrm{\bar{K}}$N)$_{I=0}$ bound state, the Binding Energy (BE) being about 27\,MeV. Based on this assumption a BE for the $\Lambda^*$-$\Lambda^*$ pair of 40\,MeV is predicted. Increasing the number of $\Lambda^*$s, the binding energy per baryon (BE/$A$) also increases as a consequence of partial restoration of chiral symmetry, whose effect is directly proportional to the baryon density. In \cite{70} it is also argued that for $\Lambda^*$ conglomerates with baryon multiplicity $A>8$ the absolute stability with respect to both the strong and the weak interactions is obtained. While the baryon density of the conglomerate increases, the mass per baryon decreases until it drops below the in-medium mass of the nucleon and the decay $\Lambda^*\rightarrow n$ is closed. This happens at baryon densities of about $3\rho_0$ (where $\rho_0$ = 0.17\,fm$^{-3}$ is the normal nuclear density) when the BE/$A$ is of about 470\,MeV (see figure 1 by \cite{17}).

Although the argument above referred on the conditions which could be achieved in the central regions of the neutron stars -- where strangeness production mechanisms (e.g. $n\to p+K^-$) become energetically favoured at large densities, leading to the formation of hyperon cores -- , the hypothesis proposed in \cite{70} could be more favourable in the cosmological field, where the possibility to have arbitrarily high values of density and temperature is not precluded. In this framework, the formation of matter with strange quark content as a partial constituent of DM could be considered more realistic. Under the hypothesis that the conditions for the formation of stable conglomerates could be set during the first phase of the Big Bang, at sufficiently high density (and temperature), we can assume that, since their formation, such conglomerates should have a very low probability to interact with baryonic matter.

In \cite{71} a critical analysis of the results obtained in \cite{70} is performed. It is argued that the model developed in \cite{70} only includes a purely attractive $\Lambda^*$-$\Lambda^*$ interaction in the $A$-body Schr{\"o}edinger equation, which leads to a divergent BE/$A$ as the baryon number $A$ increases. A relativistic mean field calculation is performed in \cite{71}, which also accounts for a repulsive $\Lambda^*$-$\Lambda^*$ interaction term. The repulsive term induces saturation of  BE/$A$ (when $A>120$, at about $2\rho_0$) to a value which is at most 100\,MeV; this is not enough to drop the mass of the $\Lambda^*$ constituents below the $\Lambda$(1116) mass. The conglomerate turns then to be instable with respect to the strong interaction decay $\Lambda^*\Lambda^*\rightarrow\Lambda\Lambda$. 

Besides the problem of the stability of SQM conglomerates with respect to the strong, weak and electromagnetic interactions, which is debated in literature and it is out of the scope of the present work, it is worth investigating the gravitational properties of SQM halos in Milky Way-size spiral galaxies to be compared with astronomical observations. In the following Sections the self-gravitational equilibrium of halos composed of SQM conglomerates is explored, considering the lower mass limit $m^*>7.46$\,GeV (proposed in \cite{70}) for the DM candidate. It is important to stress that the results obtained in Sections \ref{sdmh} and \ref{comparison} are not dependent on the nature of the DM particle, thus can be applied also to other DM candidates having mass of the same order of magnitude.

\section{Milky Way halo modelling}

\begin{figure} \centering
\includegraphics[width=0.5\textwidth]{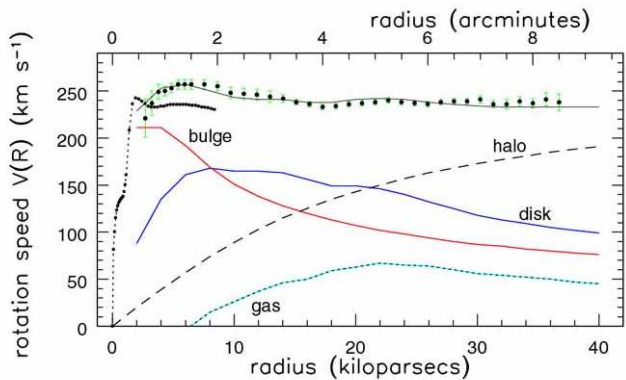}
\caption{Different components in our Galaxy (spiral) and contribution to the rotation curve (figure by K. Begeman \& Y. Sofue adapted from \cite{4a}).}
\label{fig1}
\end{figure}

Analyzing the rotation velocity curve by observative data, a flat behavior around 200\,km/s is clearly evidenced, indicating a significant difference from the expected theoric trend. The existence of a Galactic halo composed by DM, like the one first introduced by \cite{1}, can explain the observed behavior. More accurate observations have been performed and the flat behavior has been confirmed (see Fig.\,\ref{fig1}), resulting in agreement with the existence of a halo of mass $M_{\rm halo}\sim 10\,M_{\rm gal}$ and radius $R_{\rm halo}\sim 10\,R_{\rm gal}$.

Some questions arise from this preliminary analysis. What is the nature of the DM? What is the particle composition of the halo? What is the mass of these particles? The problem was widely discussed since 1970s, and the construction of DM halos models has experienced a significant development with the hypothesis of a massive neutrino (with a mass of the order of few tens of eV), generically named WIMP, as a diffuse component due to the importance of beta decay in the stellar evolution (e.g.\, \cite{14,15,16}). 

The equilibrium of such a self-gravitating halo can be solved by considering a degenerate Fermi gas of neutrinos, and using a polytropic model with $n=3/2$. The halo mass and radius are straightforward to obtain, their expressions given by
\begin{equation} \label{masspoly}
M = \frac{3}{2}\left(\frac{\pi}{2}\right)^{3/2}(2.71406)\,\frac{\hbar^3}{G^{3/2}m_{\nu}^4}\,\rho_0^{1/2},
\end{equation}
\begin{equation} \label{radpoly}
R = \frac{\left(9\pi\right)^{1/6}}{2\sqrt{2}}(3.65375)\,\frac{\hbar}{G^{1/2}m_{\nu}^{4/3}}\,\rho_0^{-1/6},
\end{equation}
where $m_{\nu}$ is the neutrino rest mass and $\rho_0$ the central density. Introducing the above conditions $M_{\rm halo}\sim 10\,M_{\rm gal}$ and $R_{\rm halo}\sim 10\,R_{\rm gal}$ implies a central density $\rho_0\sim 10^{-25}\rm{g\,cm}^{-3}$ for a neutrino rest mass $m_{\nu}\sim 10$\,eV. Moreover, combining Eqs.\,(\ref{masspoly}) and (\ref{radpoly}) leads to a simple relation between mass and radius of our Galaxy
\begin{equation} \label{msrd}
R\simeq 90\left(\frac{M}{10^{12}\,{\rm M}_{\odot}}\right)^{-1/3}\rm{kpc}\ .
\end{equation}  

A non-relativistic treatment of the halo equilibrium is clearly the most appropriate given that both the critical density $\rho_{cr}$ and the General Relativity factor $GM/Rc^2$ are small, i.e.
\begin{equation}
\rho_{cr}=\frac{m_{\nu}^4c^3}{3\pi^2\hbar^3}=7.8\cdot 10^{-17}\rm{\,g\,cm^{-3}}\gg\rho_0
\end{equation}
and
\begin{equation}
\frac{GM}{Rc^2}=4.8\cdot 10^{-7}\ll 1\ .
\end{equation}

\section{Strange dark matter halos}\label{sdmh}

\begin{figure} \centering
\includegraphics[width=0.45\textwidth]{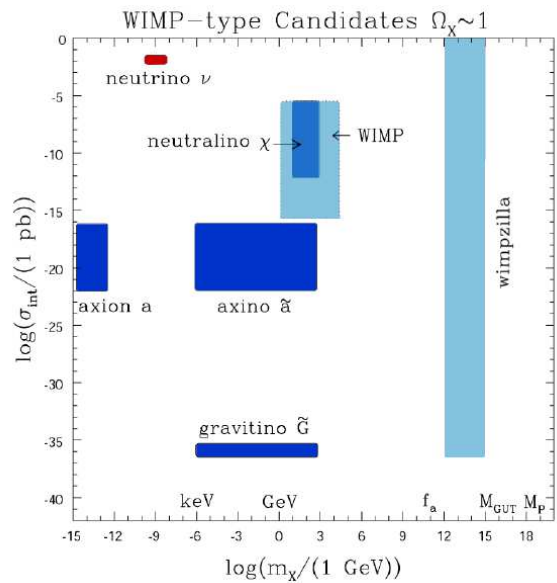}
\caption{A partial review of different DM candidates (particles only; adapted from \cite{4}).}
\label{fig2}
\end{figure}

The WIMP hypothesis is not unique in the framework of possible DM particle candidates. There are in fact a lot more candidates (fuzzy DM, hidden photons, ultra-light axions...) discussed in the literature (e.g.\,\cite{4}), with $m_{\rm DM}$ in principle anywhere between $10^{-31}$\,GeV and $10^{18}$\,GeV (see Fig.\,\ref{fig2}).

Furthermore, the possibility to have non-WIMP DM canditates should be taken into account. This alternative and fascinating hypothesis involves SMPs directly produced in the framework of the Big Bang standard model. Such a scenario arises by the simple consideration that the interaction rate between baryons and DM particles may be suppressed if DM particles are produced with large mass and consequently low number density. In fact, this rate is proportional to $n\sigma v$, with $n$ the number density, $\sigma$ the cross section and $v$ the particle velocity. Therefore, DM particles with low effective interaction rate (even for large cross sections) should evolve independently as massive Big-Bang relics, constituting a useful background in the formation of galactic halos. Among different possible candidates for DM, particles with strangeness may play a very interesting role, in particular, the SQM conglomerates discussed in Section II, with masses around $8\div 15$\,GeV. These Big Bang relics particles could form galactic halos.

Physical models for the production of free SQM conglomerate-like particles (e.g.\,``strangelets''), like the ejection of charged quark nuggets from compact objects (e.g.\,\cite{rate1}) are currently under investigation. However, a coherent scenario of conglomerate formation in the primordial Universe from pristine strange particles is presently unavailable. Therefore, we lack a detailed study of the production rates of stable SQM lumps at those cosmic epochs. Nevertheless, we can give at least a rough estimate of the upper limit to the primordial number density of such hypothetical particles. To this end, we consider an elementary number of $m\sim 1$\,GeV baryons with strangeness given by $dN_1=\rho\sigma vdt/m$ crossing a volume containing $dN_2=\rho d\mathcal{V}/m$ of such particles to form a number of conglomerates $dN_c$ of baryon number $A$ at a rate per unit volume $\dot{n}_c$. We obtain
\begin{equation} \label{prodrate}
dN_c = \frac{dN_1\cdot dN_2}{A}\ \Rightarrow\ \dot{n}_c = \frac{dN_c}{dt d\mathcal{V}} = \frac{\sigma v \rho^2}{A m^2}\,.
\end{equation}
In a very simplified scenario, we assume that the physical conditions holding when the Universe is $\sim 1$\,s old ($T\sim 10^{12}$\,K and $\rho\approx\rho_M\sim 10^5$g\,cm$^{-3}$, being $\rho_M$ the cosmological matter density) remain stationary for $\Delta t\simeq 1$\,s. Then, we derive the baryon velocity $v\sim 10^{10}$cm\,s$^{-1}$ from the relativistic equipartition $mc^2(\gamma -1)=3kT/2$, and adopt a typical baryon cross section $\sigma\sim 10^{-25}$cm$^2$ (e.g.\,\cite{rate2}). Under these assumptions, for $A\sim 10$, Eq.\,(\ref{prodrate}) yields a conglomerate density $n_c =\dot{n}_c \Delta t\sim 10^{42}$cm$^{-3}$. Actually, such a mechanism is exceedingly efficient, producing a present-day conglomerate mass density $\rho_c\sim 10^{-17}$g\,cm$^{-3}$ that is $10^{12}$ times larger than the Universe critical density $\rho_{\rm crit}\sim 10^{-29}$g\,cm$^{-3}$. It is therefore clear that an accurate estimation of the production rate of SQM conglomerates must be performed taking into account the true density of baryons with strangeness in the early Universe along with their formation and destruction rates. We address the treatment of these open issues to a forthcoming dedicated publication.

This scenario must clearly be considered as only a possible hypothesis of formation of DM, and its further investigation is needed, especially from the quantitative point of view. One of the problems is related to the expansion rate of the Universe: if cooling rate and decrease of density are in fact faster than the stabilization rate of conglomerates, the process is not implemented. Another problem is connected with the collisions among conglomerates: fluctuations of density with respect to the average value may increase the collision rate and thus create the conditions for instability of such systems. These particular conditions can also be reached in the central regions of a single galactic halo, if the central density of visible matter (galaxy) and the gravitational field are high enough to increase the probability of collision among conglomerates. During the collisions, kinetic energy can give the particles of a single conglomerate enough energy to reach a new instability, and then decay in standard model pairs that subsequently annihilate in $\gamma$-ray photons. Therefore, it is important to look into high-density regions, where the collisions are more probable, in order to obtain evidences of DM existence through the indirect detection of $\gamma$-rays from DM self-interaction.

In order to calculate self-gravitating equilibrium configurations of DM halos, we explore the possibility of having halos composed by stable SQM conglomerates. Despite the high mass density of the internal structure constituting each conglomerate ($\rho >10^{15}\rm{\,g\,cm}^{-3}$ is the lower limit obtained from the model proposed in \cite{17}), this value is not relevant in the modeling of galactic halos, where such conglomerates interact only gravitationally, and the halo mean density is of the order of $10^{-26}\rm{g\,cm}^{-3}$. 

\begin{figure} 
\centering
\includegraphics[width=0.55\textwidth]{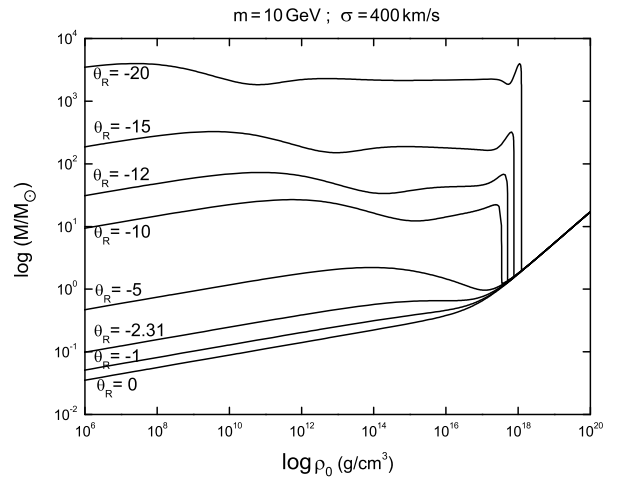}
\caption{Mass $M$ of the equilibrium configurations in function of central density $\rho_0$, at different values of $\theta_R$. A phase transition from non-quantum to fully degenerate configurations at densities around $10^{18}\rm{g\,cm}^{-3}$ is clearly visible.}
\label{fig3}
\end{figure}
\begin{figure}
\centering
\includegraphics[width=0.55\textwidth]{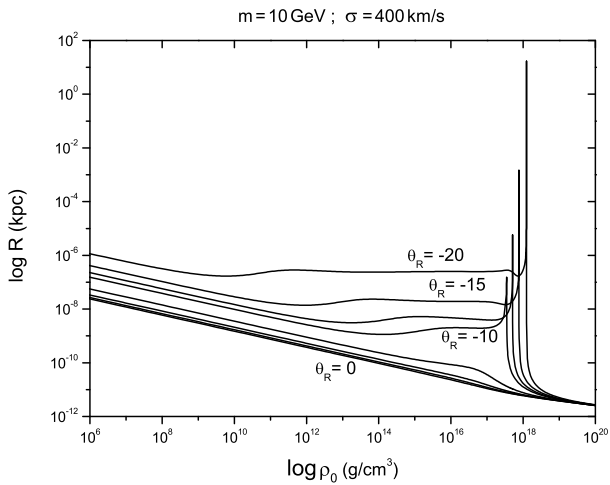}
\caption{Radius $R$ of the equilibrium configurations in function of central density $\rho_0$, at different values of $\theta_R$. A phase transition from non-quantum to fully degenerate configurations at densities around $10^{18}\rm{g\,cm}^{-3}$ is clearly visible.}
\label{fig4}
\end{figure}

First, we consider a semi-degenerate gas of particles with a rest mass $m^*=8\div 15$\,GeV. We look for halos with mass $M\sim 10^{12}\,{\rm M}_{\odot}$ and radius $R\sim 100$\,kpc, with a mean density $\bar\rho$ of the order of $10^{-26}\rm{g\,cm}^{-3}$. In particular, for $m^*=10$\,GeV, we obtain
\begin{equation}
\rho_{cr}=\frac{m^{*4}c^3}{3\pi^2\hbar^3}=7.8\cdot 10^{19}\rm{\,g\,cm}^{-3}\gg\,\bar\rho\ ,
\end{equation}
and
\begin{equation}
\frac{GM}{Rc^2}=4.8\cdot 10^{-7}\ll 1\ .
\end{equation}
This demonstrates that also strange DM halos are non-relativistic and Newtonian.

For the equilibrium configuration, we consider a semi-degenerate Fermi distribution function with a cutoff in energy given by the following expression \cite{18}
\begin{equation} \label{semidegdistr}
\begin{cases}
f(\varepsilon) = \dfrac{g}{h^3}\left[
\dfrac{1-e^{(\varepsilon -\varepsilon_c)/kT}}{e^{(\varepsilon -\mu)/kT}+1}
\right] & \mbox{ for }\varepsilon\leq \varepsilon_c \\
f(\varepsilon) = 0 & \mbox{ for }\varepsilon >\varepsilon_c\ ,
\end{cases}
\end{equation}
where $\varepsilon_c =m(\varphi_R-\varphi)$ is the cutoff energy, $\varphi$ is the gravitational potential, 
$\mu$ is the chemical potential and $g=2s+1$ is the multiplicity of quantum states. The mass density $\rho$ is given by
\begin{equation}
\rho\ =\ m \int f(\varepsilon)\ d^3q \ .
\end{equation}

For the gravitational equilibrium, we use the Poisson equation
\begin{equation} \label{poisson}
{\frac{1}{r^2}}{\frac{d}{dr}}\left({r^2\frac{d\varphi}{dr}}\right)=4\pi G\rho\ ,
\end{equation}
with $\varphi '(0)=0$ and $\varphi(0)=\varphi_0$.

By integrating Eq.\,(\ref{poisson}), we obtain different equilibrium configurations at different values of $W_0$ and $\theta_R$, where $W_0$ is the value of $W=\varepsilon_c/kT$ at the center of the configuration and $\theta_R$ is the
value of $\theta=\mu/kT$ at the border of the configuration. These quantities are related by the condition $\theta_R =\theta -W\leq 0$ arising from the inequality $(\partial U/\partial N)_{S,V}\leq\epsilon_c\,$, being that the energy variation due to a single particle (at constant entropy $S$ and volume $V$) cannot be larger than the maximum energy that a particle can have \cite{19,20}. The solutions also depend on $m$ (mass of the particle) and $\sigma$ (surface velocity dispersion) through scaling laws. The results are summarized in diagrams of $M$ versus $\rho_0$ and $R$ versus $\rho_0$ for $m=10$\,GeV and $\sigma=400$\,km\,s$^{-1}$ (see Fig.\,\ref{fig3} and Fig.\,\ref{fig4}, respectively).

It is clear that the particle mass value $m=10$\,GeV does not allow to obtain the expected values of central density, mass and radius for a galactic halo. In fact, we have $\rho_0\propto\sigma^3 m^4$, $M\propto\sigma^{3/2}m^{-2}$ and $R\propto\sigma^{-1/2}m^{-2}$. This results in too large densities, and too small masses and radii, implying that the semi-degenerate regime is not appropriate to describe strange DM halos. We need much more negative values of $\theta_R$, typical of a classical regime well described by the Boltzmann (King) distribution function with cutoff in energy. In conclusion, strange DM halos are non-relativistic, Newtonian and do not follow quantum statistics; therefore any speculation on spin of the particles is neither decisive nor pertinent for the description of this system.

\begin{figure}
\centering
\includegraphics[width=0.55\textwidth]{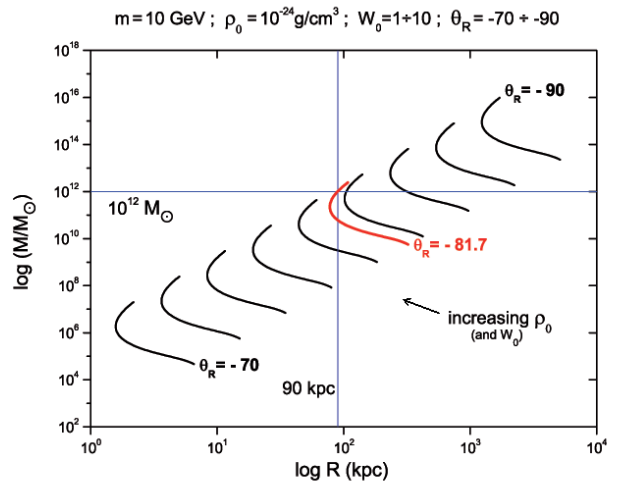}
\caption{Identification of the value of $\theta_R$ compatible with the required values of mass and radius for spiral-galaxy halos. The position of a MW-sized halo (\textit{blue axes}) is highlighted, along with the relevant value of $\theta_R$ (\textit{red curve}).}
\label{fig5}
\end{figure}

\begin{figure}
\centering
\includegraphics[width=0.55\textwidth]{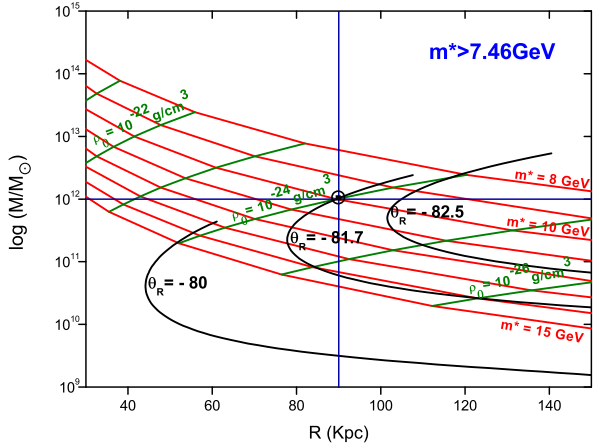}
\caption{Solutions for MW-sized DM halos obtained through scaling laws between halo physical parameters and conglomerate masses. The relations between halo mass and size for different conglomerate masses (\textit{red curves}), DM central densities (\textit{green curves}) and concentration parameters (\textit{black curves}) are plotted. The position of the MW-sized halo for $m^*=10$\,GeV (\textit{open circle with dot}) is also marked.}
\label{fig6}
\end{figure}

In order to obtain halos with appropriate densities, masses and radii, we calculate equilibrium configurations at fixed central density $(\rho_0=10^{-24}\rm{g\,cm}^{-3})$ and particle mass ($m=10$\,GeV), while increasing the value of $-\theta_R$ until we reach $M\sim 10^{12}\,{\rm M}_{\odot}$ and $R\sim 100$\,kpc (see Fig.\,\ref{fig5}). We compute solutions in the range $W_0=1\div 10$ (for globular clusters, the most significant values are between 4 and 8; for galactic halos we expect even less). In this regime, the dependence on 
$\theta_R$ becomes a scaling law. It is possible to make a tuning by varying the central density $\rho_0$ and the concentration parameter $\theta_R$ in order to match the required values in $M$ and $R$, also at different values of $W_0$. The obtained results for $m=10$ GeV and $\rho_0=10^{-24}\rm{g\,cm}^{-3}$ are very satisfying: we obtain $\theta_R=-81.7$ and $W_0=1.8$, implying a halo mass $M=9.98\cdot 10^{11}\,{\rm M}_{\odot}$, a halo radius $R=89.41$\,kpc, a mean halo density ${\bar\rho}=3M/4\pi R^3=2.16\cdot 10^{-26}\rm{g\,cm}^{-3}$ and a velocity dispersion $\sigma=405$\,km\,s$^{-1}$. The other solutions can be obtained from scaling laws involving the total mass $M$ and the radius $R$. We obtain
\begin{equation} \label{massfin}
M = 9.98\cdot 10^{11}\,\left(\frac{\rho_0}{10^{-24}\rm{g\,cm}^{-3}}\right)^{1/2}\left({\frac{m^*}{10\rm{\, GeV}}}\right)^{-4}\rm{\,M}_{\odot}\ ,
\end{equation}
\begin{equation} \label{radfin}
R = 89.41\,\left(\frac{\rho_0}{10^{-24}\rm{g\,cm}^{-3}}\right)^{-1/6}\left(\frac{m^*}{10\rm{\,GeV}}\right)^{-4/3}\rm{\,kpc}\ .
\end{equation}
Here we indicated the particle mass $m$ as the conglomerate mass $m^*$.

Analogously, we could express the Eqs.\,(\ref{massfin}) and (\ref{radfin}) in terms of $\sigma$ and $m^*$ or $\sigma$ and $\rho_0$. In these cases we get $M\propto\sigma^{3/2}m^{*-2}$ and $R\propto\sigma^{-1/2}m^{*-2}$, or $M\propto\sigma^{3}\rho_0^{-1/2}$ and $R\propto\sigma\rho_0^{-1/2}$. It is also interesting to express the surface velocity dispersion $\sigma$ as a function of the other two parameters, the conglomerate mass $m^*$ and central density $\rho_0$; we get
$$\sigma = 405\,\left(\frac{\rho_0}{10^{-24}\rm{g\,cm}^{-3}}\right)^{1/3}\left({\frac{m^*}{10\rm{\, GeV}}}\right)^{-4/3}\rm{km\,s^{-1}}\ .$$
These results are summarized in Fig.\,\ref{fig6}.

\section{Comparison with dwarf galaxy halo properties}\label{comparison}

The theoretical scenario presented in Section \ref{sdmh}, though fascinating, is deeply related to the existence of unobserved SQM conglomerates; furthermore, the derived physical parameters of the DM halo hold in principle only for MW-sized spiral galaxies. In this Section, we address such issues by taking advantage of  DM halos around different classes of galaxies sharing similar structural properties. In particular, we take advantage of the scenario of massive DM particles annihilating or decaying into SM products, among which $\gamma$-rays \cite{72}, which is being currently explored with deep observations of DM halos at several $\gamma$-ray facilities for DM particle masses from the GeV to the TeV range (up to $\sim 100$\,TeV). In the case of SQM conglomerates, the mass increases with the baryon multiplicity $A$: any value for the conglomerate mass is thus allowed if the stability conditions are fulfilled. In particular, in the model by \cite{70} the stability is achieved for $A>8$, which corresponds to $m^*>7.46$ GeV. Therefore, SQM conglomerates naturally lie in the GeV-to-TeV particle mass range.

\begin{table*}
\caption{Best-fit DM halo parameters obtained with {\scriptsize CLUMPY} for the nine dSphs analyzed in this study. In the dSph type column, ``cls'' stands for ``classical'' and ``uft'' for ``ultra-faint''. Distances $d_\odot$ are taken from \cite{21,92}.}
\medskip
\centering
\resizebox{\textwidth}{!}{
\renewcommand{\arraystretch}{1.5}
\begin{tabular}{lccccccc}
\hline
\hline\\
Name & Type & $d_\odot$(kpc) & $\rho_s$($10^{-26}$g\,cm$^{-3}$) & $r_s$(kpc) & $\alpha$ & $R_{\rm vir}$(kpc) & $M_{\rm DM}($<$R_{\rm vir})/10^8$\,M$_{\odot}$\\\\
\hline
&&&&\\
Carina (Car) & cls & $105 \pm 6$ & $28^{+104}_{-19}$ & $1.8^{+6.5}_{-1.2}$ & $0.31^{+0.54}_{-0.15}$ & $5.0^{+6.8}_{-3.7}$ & $3.0^{+6.1}_{-1.1}$\\
Coma Berenices (CBe) & uft & $44 \pm 4$ & $250^{+370}_{-190}$ & $1.0^{+4.6}_{-0.8}$ & $0.64^{+0.25}_{-0.29}$ & $4.2^{+18.7}_{-1.5}$ & $7.5^{+159.0}_{-0.4}$\\
Canes Venatici I (CVn\,I) & uft & $218 \pm 10$ & $17^{+45}_{-10}$ & $1.9^{+2.8}_{-1.1}$ & $0.32^{+0.32}_{-0.13}$ & $12.2^{+15.4}_{-8.2}$ & $7.3^{+12.3}_{-2.4}$\\
Draco I (Dra\,I) & cls & $76 \pm 6$ & $580^{+450}_{-200}$ & $0.279^{+0.046}_{-0.084}$ & $0.41^{+0.24}_{-0.21}$ & $28^{+31}_{-21}$ & $740^{+1080}_{-350}$\\
Reticulum II (Ret\,II) & uft & $32 \pm 2$ & $440^{+1480}_{-340}$ & $0.4^{+2.3}_{-0.3}$ & $0.56^{+0.29}_{-0.26}$ & $1.7^{+7.6}_{-0.6}$ & $1.0^{+13.4}_{-0.1}$\\
Sculptor (Scl) & cls & $86 \pm 6$ & $136^{+70}_{-93}$ & $0.69^{+0.43}_{-0.23}$ & $0.28^{+0.51}_{-0.11}$ & $8.5^{+11.7}_{-7.5}$ & $19^{+48}_{-14}$\\
Segue 1 (Seg\,1) & uft & $23 \pm 2$ & $11^{+157}_{-9}$ & $0.3^{+4.0}_{-0.2}$ & $0.53^{+0.33}_{-0.27}$ & $0.3^{+2.2}_{-0.1}$ & $0.0041^{+0.0710}_{-0.0003}$\\
Tucana II (Tuc\,II) & uft & $58 \pm 5$ & $26^{+106}_{-20}$ & $1.4^{+3.8}_{-1.0}$ & $0.60^{+0.31}_{-0.29}$ & $2.6^{+6.5}_{-1.1}$ & $1.5^{+5.2}_{-0.2}$\\
Ursa Minor (UMi) & cls & $76 \pm 3$ & $10^{+8}_{-6}$ & $4.3^{+2.7}_{-1.0}$ & $0.28^{+0.15}_{-0.12}$ & $10.4^{+10.5}_{-8.6}$ & $42^{+42}_{-27}$\\
&&&&\\
\hline
\end{tabular}
}
\label{tab1}
\end{table*}

Here, we show how the average density of a strange DM halo is common also to halos of different size like those surrounding the dwarf spheroidal galaxies (dSphs), probably the most DM dominated objects in the local Universe, and that such halos can be obtained by scaling down the typical masses and radii for halos around normal galaxies. In 2015, the {\itshape Fermi}-LAT $\gamma$-ray observatory discovered a $\gamma$-ray excess between 
$\sim 3$ and $\sim 10$\,GeV in the direction of the dSph Reticulum II (Ret\,II; \cite{21})\footnote[1]{Another tantalizing detection of a $\gamma$-ray excess is reported for the Galactic center by \cite{22}; however, its interpretation as a product of DM self-interaction with $m_\chi\sim 45$\,GeV is still controversial.}. Such an excess was compatible with a flux due to annihilation of DM particles with mass $m_\chi\sim 25$\,GeV at 
$3\sigma$ confidence level (see figures 1 and 3 by \cite{13}).

In order to highlight the structural and physical similarities between dSph and MW DM halos, we derive the amount of DM in a sample of selected dSphs by analyzing the kinematics of their member stars. To this end, we apply the Markov-Chain Monte Carlo (MCMC) Jeans analysis integrated in the {\scriptsize CLUMPY}\footnote[2]{Available at {\ttfamily http://clumpy.gitlab.io/CLUMPY/index.html}.} software \cite{26,27,75} and described in \cite{30} to the dSphs analyzed by \cite{73}, with the inclusion of Carina (Car), Tucana II (Tuc\,II) and Ursa Minor (UMi) and the removal of Triangulum II (Tri\,II; see \cite{76,77}). We present the list of selected targets in Tab.\,\ref{tab1}.

We refer to \cite{29,30} for a detailed description of the spherical Jeans analysis. Here, we simply recall that the integration of the moments of the phase-space distribution function for a steady-state, spherically symmetric and negligibly rotating collisionless system yields the second-order Jeans equation \cite{25}
\begingroup\makeatletter\def\f@size{8}\check@mathfonts
\begin{equation}\label{distfunc}
\frac{1}{n(r)}\left[\frac{d}{dr}\left(n\bar{v}_r^2\right)\right]+2\frac{\beta_{\rm ani}(r)}{r}\bar{v}_r^2(r)= -\frac{4\pi G}{r^2}\int_0^r\rho_{\rm DM}(s)s^2 ds\ .
\end{equation}
\endgroup
Here, $n(r)$, $\bar{v}_r^2(r)$ and $\beta_{\rm ani}(r)$ are the stellar number density, velocity dispersion and velocity anisotropy respectively. For the case of dSphs, the solution to Eq.\,(\ref{distfunc}) relates the internal proper motions of stars to the amount of DM in the dSph halo, although only line-of-sight observables like the projected radius $R$, the surface brightness $\Sigma(R)$ and the projected stellar velocity dispersion $\sigma_p(R)$ can be directly compared with data.

In order to determine the parameters that best reproduce the observed properties of the selected dSphs, we run a set of $10^5$ MCMC simulations with {\scriptsize CLUMPY} on the member stars of each target, according to the prescriptions for an unbinned analysis described in \cite{29,30} and assuming an Einasto profile \cite{31} for the DM distribution given by
\begin{equation}\label{einasto}
\rho(r) = \rho_s\,e^{-\frac{2}{\alpha}\left[\left(r/r_s\right)^\alpha -1\right]}\ .
\end{equation}
We selected this DM profile over other possible choices, such as the Zhao-Hernquist (ZH; \cite{32,33}) or the (cored) Burkert profiles \cite{89}, since $N$-body simulations of the Local Volume have established that non-singular cuspy profiles are well suited to universally describe DM subhalos gravitationally bound to MW-sized galaxies (e.g.\,\cite{90}). In addition, \cite{30} find no or negligible differences in the shapes of DM densities obtained by fitting either Einasto or ZH profiles.

We preventively calculate the stellar number density $n(r)$ by fitting a 3D ZH profile to publicly available 2D photometric data of our targets \cite{21,78,79,80}, and the resulting parameters are used as a fixed input for {\scriptsize CLUMPY}. We take the stellar-kinematics data from the most up-to-date literature for each target \cite{28,82,83,84,85,86,87}, and estimate the membership probability $P$ of the member stars as follows:
\begin{itemize}
\item for the ``classical'' dSphs Car, Draco I (Dra\,I), Sculptor (Scl) and UMi, and the ``ultra-faint'' Segue 1 (Seg\,1), we apply an estimation-of-membership (EM) algorithm \cite{35} to their member candidates, keeping in the Jeans analysis only those for which $P \geq 0.95$;
\item for the remaining ``ultra-faint'' dSphs Coma Berenices (CBe), Canes Venatici I (CVn\,I), Ret\,II and 
Tuc\,II, we associate binary (0/1) memberships taken from the literature to each member candidate.
\end{itemize}
Finally, we adopt the functional form for the stellar velocity anisotropy $\beta_{\rm ani}(r)$ by \cite{88}.

\begin{figure*}
\centering
\includegraphics[width=.99\textwidth]{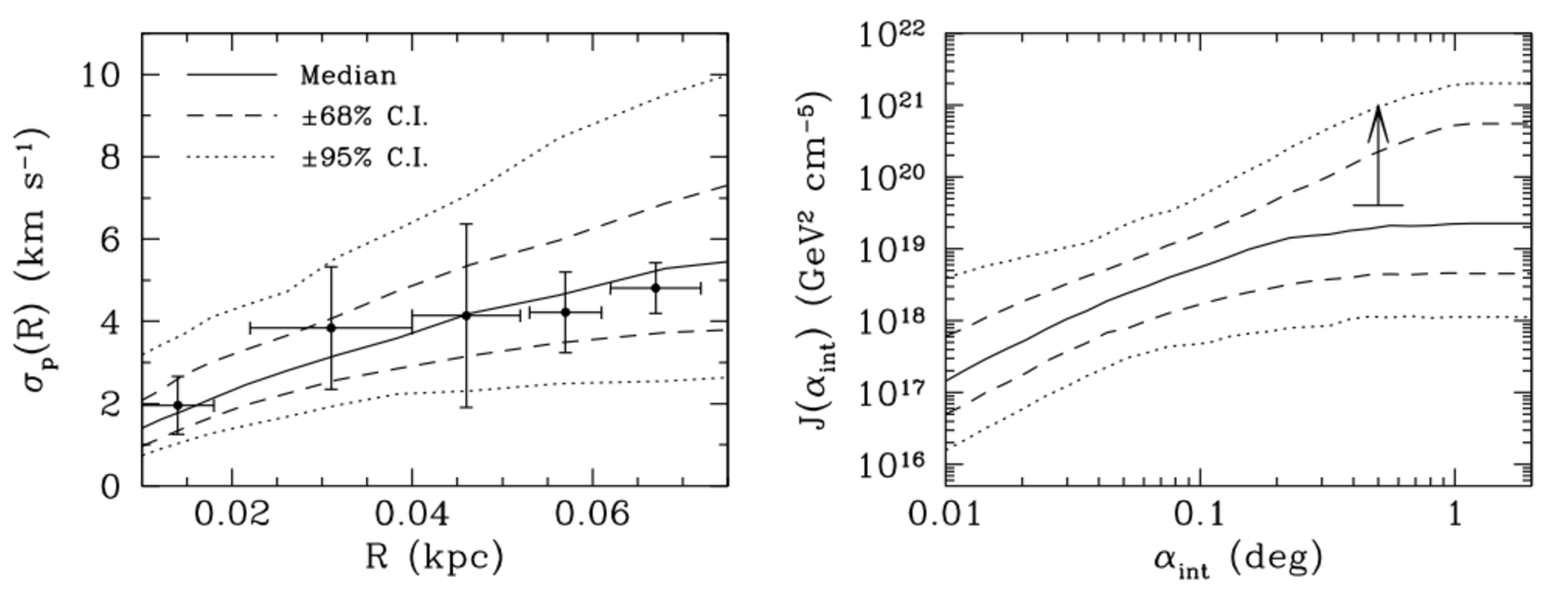}
\caption{{\itshape Left panel:} Ret\,II stellar velocity dispersion as a function of the projected radius from the dSph center obtained from the MCMC Jeans analysis. The median profile (\textit{solid line}) is shown together with the corresponding confidence intervals at 68\% (\textit{dashed lines}) and 95\% confidence level (\textit{dotted lines}). For comparison, the binned measurements from \cite{86} (\textit{dots}) are overplotted together with their $1\sigma$ errors computed over the identified member stars. {\itshape Right panel:} Ret\,II astrophysical factor for DM annihilation as a function of the instrumental integration angle. The median profile (\textit{solid line}) is shown together with the corresponding confidence intervals at 68\% (\textit{dashed lines}) and 95\% confidence level (\textit{dotted lines}). The lower limit on $J(\alpha_{int})$ derived by \cite{13} is reported as a visual confirmation of the goodness of the MCMC calculations.}
\label{fig7}
\end{figure*}

In this way, we are able to fit Eq.\,(\ref{distfunc}) to the input data in order to obtain the posterior distributions of the 7 free parameters (4 for the velocity anisotropy profile and 3 for the DM density profile). We list the best-fit parameters for the DM density profile of each analyzed dSph in Tab.\,\ref{tab1}, along with the corresponding virial radius $R_{\rm vir}$ computed as in \cite{30} (see their equation 18; see also \cite{25}) taking the dSph distances from \cite{21,92}, and the enclosed DM mass $M_{\rm DM}($<$R_{\rm vir})$. We also verify that our MCMC simulations are self-consistent by reproducing the projected stellar velocity dispersion profiles $\sigma_p(R)$ and the astrophysical factors $J(\Delta\Omega)$ for DM annihilation of the analyzed dSphs, expressed by \cite{91}
\begin{equation}\label{jfact}
J(\Delta\Omega) = \int_{\Delta\Omega} d\Omega \int_{\rm l.o.s.} \rho_{\rm DM}^2(\ell, \Omega) d\ell\ .
\end{equation}
Such profiles are compatible within errors with those obtained by \cite{24,30}, and their analysis will be demanded to a forthcoming publication (Doro, Morselli, Rodr{\'i}guez-Fern{\'a}ndez, Saturni et al.; in prep.). As an example, we show the profile of $\sigma_p$ for Ret\,II in Fig.\,\ref{fig7}, along with the profile of $J$ as a function of the integration angle $\alpha_{\rm int}$ to be compared with the lower limit inferred by \cite{13} from the $\gamma$-ray excess.

It is to be noted that, due to the steady-state spherical symmetry of dSph DM halos assumed by the Jeans analysis implemented in {\scriptsize CLUMPY}, the impact of stellar feedback, triaxiality and tidal interactions on the dynamical status of such halos remains unknown \cite{29,36}. Nevertheless, the average density $\bar{\rho}$ of dSph halos estimated from virial radii and masses listed in Tab.\,\ref{tab1} ranges between $6.57 \cdot 10^{-27}$g\,cm$^{-3}$ and $3.13\cdot 10^{-25}$g\,cm$^{-3}$, with a mean value $\langle\bar{\rho}\rangle=(1.26\pm 0.93)\cdot 10^{-25}$g\,cm$^{-3}$ in agreement at a 95\% confidence level with the value for the MW derived in Section \ref{sdmh}. The fact that different halos associated to morphologically very different galaxies, dominated in different way by the DM component -- in fact, the dSph mass-luminosity ratios are very high with respect to those of common galaxies -- have mean densities compatible among them, encourages the continuation of studies on their common origin. Therefore, if the hypothesis claiming SQM conglomerates to constitute galaxy halos is valid, the DM particles composing such halos must have formed immediately after the Big Bang, when the energy density was in the correct range to allow their production and, subsequently, ensure their stability. Decoupled from ordinary matter, they would have been able to gravitationally aggregate, forming the potential wells where proto-galaxies began to collapse.

\begin{figure}[!] 
\centering
\includegraphics[width=0.99\columnwidth]{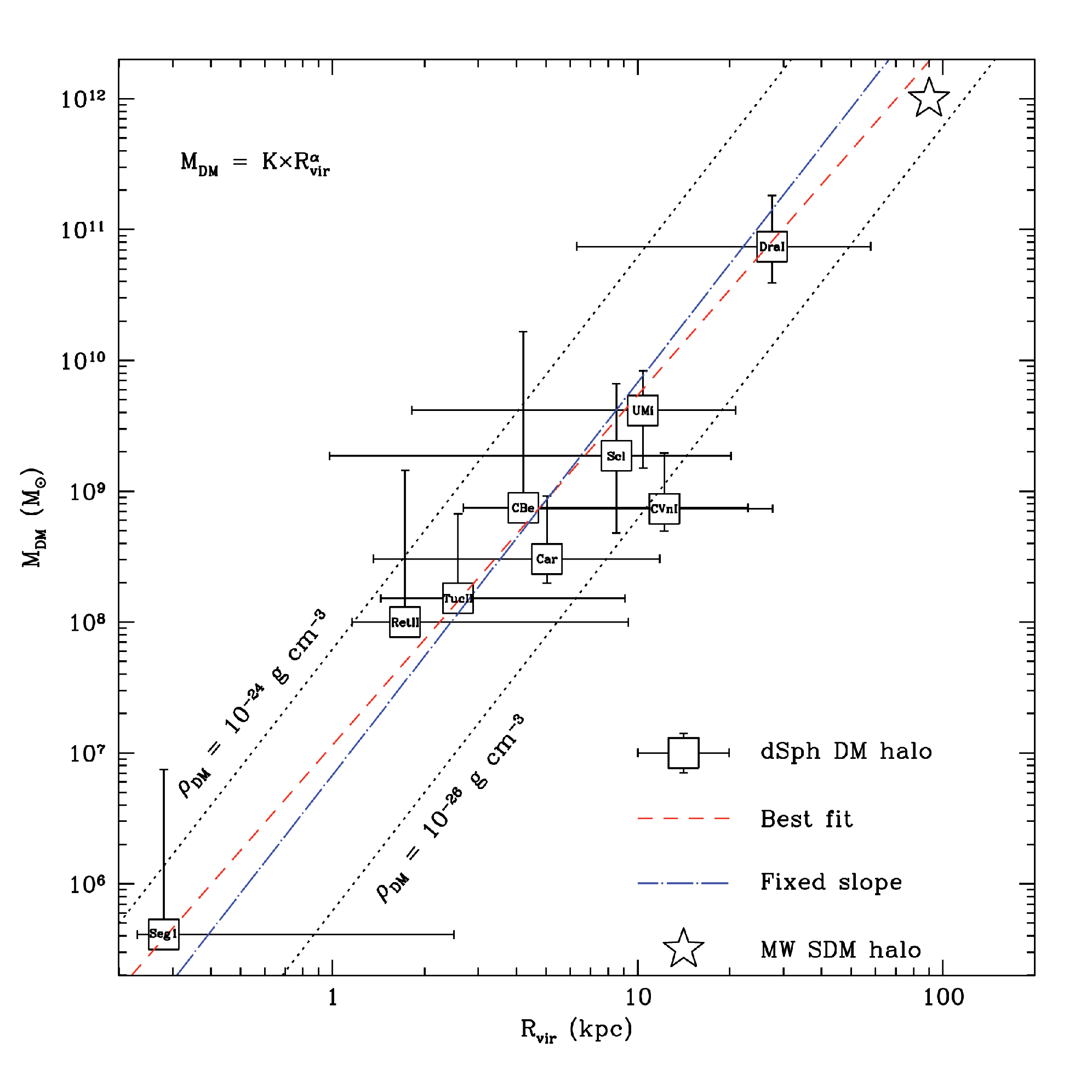}
\caption{Scaling relation for nine dSph DM halo parameters obtained from the Jeans analysis performed with the CLUMPY software (\textit{open squares}) and MW parameters for the theoretical DM halo constructed with SMPs (\textit{open star}). The errors at 68\% confidence level are associated to the measurements of halo masses and virial radii. For comparison, the relation $M_{\rm DM}\propto R_{\rm vir}^3$ (\textit{dashed line}) -- corresponding to the average density of $\sim 1.26 \cdot 10^{-25}$g\,cm$^{-3}$ -- is reported, along with the same relation scaled at $10^{-24}$ and $10^{-26}$g\,cm$^{-3}$ (\textit{dotted lines}).}
\label{fig8}
\end{figure}

In Fig.\,\ref{fig8} we show the scaling relation between the dSph halo parameters obtained from the MCMC spherical Jeans analysis with {\scriptsize CLUMPY} and the MW parameters for the theoretical SMP halo, using the numerical data summarized in Tab.\,\ref{tab1}. A visual inspection already reveals the correlation between $\log{M_{\rm DM}}$ and $\log{R_{\rm vir}}$, which is confirmed by a correlation coefficient $r = 0.98$ and an associated null-hypothesis probability $p(<r) = 1.05 \cdot 10^{-6}$ (e.g.\,\cite{93}). In order to quantitatively compare dSph and MW halos, we perform a logarithmic bisector fit \cite{94} restricted to the dSph data weighted for their uncertainties along both axes, adopting either no constraints on the free parameters or with logarithmic slope (i.e. power-law index $a$) fixed to 3. The best-fit relation is given by
\begin{equation}\label{bestfit}
\log{\left[\frac{M_{\rm DM}(\mbox{<}R_{\rm vir})}{10^9\,{\rm M}_{\odot}}
\right]}=a\log{\left(\frac{R_{\rm vir}}{10\,{\rm kpc}}\right)}+b\ ,
\end{equation} 
with $a=2.67\pm 0.57$ and $b=0.73\pm 0.50$ ($\chi^2/n_{\rm d.o.f.} = 9.6/7$). The fit of the fixed-slope relation yields instead $b=0.83\pm 0.30$ ($\chi^2/n_{\rm d.o.f.} = 8.8/8$).

It is clear that, with such values of the statistical goodness of fit at hand, we cannot strongly prefer a relation over the other: in fact, according to an $F$-test, the best-fit relation represents a statistical improvement at 60\% confidence level only when compared to the fit with fixed logarithmic slope. In addition, the values of $a$ and $b$ are compatible within errors in both models. Therefore, we can conclude that DM halos around galaxies with different morphology and stellar content can be approximated with spheres of mean density $\sim 1.26\cdot 10^{-25}$g\,cm$^{-3}$ over a range of almost three orders of magnitude in virial radius and six orders of magnitude in enclosed DM mass.

\section{Conclusions}

In this paper, we proposed a possible scenario for DM origin in the Universe based on conglomerates made of strange quark matter. These conglomerates form in the very early phases after the Big Bang, when the conditions of extreme density and temperature may favor the aggregation of strange baryonic matter in stable structures that interact only gravitationally with ordinary matter. Subsequently, when the Universe expands and cools down, the conglomerates formed in this way settle into galactic halos as ``relic'' DM.

We also showed how the assumption of conglomerates with mass of $\sim 10$\,GeV can lead to a good reproduction of the physical properties (mass, radius, concentration) of a typical MW-sized DM halo. Performing a Jeans analysis on the kinematical properties of nine dSphs, we also showed how the average DM density in halos of very different size is approximately maintained, hinting for a common origin of both families of structures.

As important remark, we recommend to adopt some caution when considering the results presented here. In fact, the proposed scenario for DM composed by conglomerates of particles with strangeness is still tentative. At present, no quantitative models able to compute the stability and formation rate of strange conglomerates exist. Consequently, we stress that the existence and stability of the conglomerates is not yet definitely established (e.g.\,\cite{71}), even if, for what concerning the results, this fact does not impact on the outcome and the conclusions, being conglomerates here assumed as a possible example of stable particle with mass of $\sim 10$\,GeV. Furthermore, while for a detailed calculation of the formation rate of such particles we address to a forthcoming paper, nevertheless we have shown as even the non-extreme physical conditions holding in the first seconds of the early Universe may suffice in producing amounts of conglomerates large enough to significantly contribute to the DM total mass. 

Finally, the detection of $\gamma$-ray signals from DM halos is still controversial, and future observations with next-generation $\gamma$-ray telescopes (e.g.\,\cite{95,96}) are needed in order to eventually achieve the detection of $\gamma$-rays produced by DM annihilation or decay in astrophysical sources.

\section*{Acknowledgements}
\noindent
Part of this work was supported by EU STRONG-2020 project (grant agreement No.\,824093).

\end{document}